\documentstyle[11pt]{article}
\begin{document}
\baselineskip 22pt
\rightline{CU-TP-734}
\rightline{hep-th/9601097}
\vskip 2cm
\centerline{\Large\bf Electromagnetic Duality and $SU(3)$ Monopoles}
\vskip 1cm
\centerline{\large\it Kimyeong Lee,\footnote{electronic mail: 
klee@phys.columbia.edu}  Erick J. Weinberg,\footnote{electronic mail:
ejw@phys.columbia.edu} and  Piljin Yi\footnote{electronic mail:
piljin@phys.columbia.edu}}
\vskip 2mm
\centerline{Physics Department, Columbia University, New York, NY 10027}

\vskip 2cm
\centerline{\bf ABSTRACT}
\vskip 1cm

\begin{quote}
We consider the low-energy
dynamics of a pair of distinct fundamental monopoles 
that arise in the $N=4$ supersymmetric $SU(3)$ Yang-Mills theory broken to 
$U(1)\times U(1)$. Both the long distance interactions and the short distance 
behavior indicate that the moduli space is $R^3\times(R^1\times {\cal M}_0)
/Z$ where ${\cal M}_0$ is the smooth Taub-NUT manifold, and we confirm this 
rigorously. By examining harmonic forms on the moduli space, we find a 
threshold bound state of two monopoles with a tower of BPS dyonic states
built on it, as required by Montonen-Olive duality. We also present 
a conjecture for the metric of the moduli space for any number of distinct 
fundamental monopoles for an arbitrary  gauge group.
\end{quote}

\newpage

\setcounter{footnote}{0}
The longstanding conjecture \cite{dual} that certain supersymmetric field 
theories possess an electric-magnetic duality has received further support 
in recent years from the calculation of dyon spectra. In theories with
spontaneously broken $SU(2)$ symmetry, calculations based on a
knowledge of the two-monopole moduli space have verified that the
spectrum of particle states carrying two units of magnetic charge is
consistent with the duality conjecture both in an $N=4$ 
\cite{sen} and an $N=2$ theory \cite{sethi}.  The extension of these results 
to higher magnetic charge has also been discussed \cite{porrati}.

In this note, we consider the extension of this analysis to larger
groups.  We concentrate in particular on the case of an $N=4$ supersymmetric
Yang-Mills theory with an arbitrary gauge $G$ of rank $n$ that is broken by 
the adjoint representation Higgs field to its Cartan subgroup, $(U(1))^n$.  
Even in the purely magnetic sector, such theories present
a challenge for the duality conjecture.  Their particle spectrum
contains ${\rm dim}G-n$ electrically charged massive vector mesons, 
one for each root of the Lie algebra.  The most
obvious candidates for the dual states are those corresponding 
to the spherically symmetric classical  solutions
obtained by imbedding the $SU(2)$ monopole and antimonopole using the
subgroups defined by the various roots.  However, counting of the zero modes
about these solutions suggests that only $n$ of these,
corresponding to a special set of simple roots, should be understood
as fundamental monopoles and that the remainder should be interpreted as
superpositions of these fundamental solitons \cite{erick}. 

As a specific example, consider an $SU(3)$ theory broken to $U(1)\times
U(1)$. If we denote the Higgs expectation value in a unitary gauge by a vector
${\Phi_i}$ in the Cartan subalgebra with generators $H_i$, there is a
unique pair of simple roots $\alpha$ and $\beta$ such that $\alpha\cdot \Phi$
and  $\beta\cdot \Phi$ are both positive.  There is  also a third positive
root,  $\gamma=\alpha+\beta$.  Associated with these  roots
are electrically charged vector mesons of masses $ \alpha\cdot\Phi$, 
$\beta\cdot \Phi$, and $\gamma\cdot\Phi$.  On the magnetic side,
there are spherically symmetric classical solutions with magnetic charges
$\alpha^*$, $\beta^*$, and $\gamma^*$ (where  $\alpha^*\equiv
\alpha/\alpha^2$, etc.).  The first two of these each have four zero modes,
associated with spatial translations and a $U(1)$ rotation.  In contrast, the
third, and indeed any BPS \cite{bps}
solution with magnetic charge $\gamma^*$, has eight
zero modes and seems to be just one member of a large family of
two-particle solutions containing one $\alpha^*$ and one $\beta^*$ monopole.  

This suggests that the dual counterpart of the $\gamma$ vector boson should
be a threshhold bound state of an $\alpha^*$ and a $\beta^*$ monopole.   A
similar phenomenon was recently found in an $N=2$ $SU(2)$ theory 
\cite{sethi}, 
where the BPS monopole is actually dual to a quark in the fundamental 
representation. 
The dual of the massive charged vector boson is a bound state found by
quantizing the low energy dynamics of two monopoles as a supersymmetric
sigma model whose target space is the two-monopole moduli space.

The obvious difficulty in pursuing this approach in the $SU(3)$ case has
been that the moduli space was previously known only for two monopoles of
identical charge \cite{atiyah}.  However, it turns out not to be too 
difficult to determine the moduli space for a pair of monopoles with charges
$\alpha^*$ and $\beta^*$.  We will present here a schematic derivation; more
details will be given in Ref.~\cite{klee}.

Since we can factor out the center of mass coordinates and a $U(1)$ phase, 
the moduli space is isometrically decomposable in the form \cite{klee},
\begin{equation}
{\cal M}=R^3\times \frac{R^1\times {\cal M}_0}{\cal G}
\end{equation}
The $R^1$ is parametrized by a coordinate $\chi$ whose conjugate momentum
$P_\chi$ we will later identify with a ``total'' electric charge. Unless the 
mass-ratio of the two monopoles is rational, $\chi$ is not in general a
periodic coordinate. The identification by $\cal G$ is closely tied to the
quantization of dyons, and we will return to the matter of this infinite
discrete group $\cal G$ later.

A natural starting point for determining ${\cal M}_0 $ is
to study the interactions of two widely separated dyons.  For the
$SU(2)$ theory, Manton showed that these forces imply that the
metric of the relative coordinate moduli space asymptotically approached
a Taub-NUT metric with negative length parameter \cite{manton}. 
This metric has a short distance singularity that renders it unacceptable
as the moduli space. But even apart from this singularity,
symmetry arguments alone show that the two metric cannot be identical.  
The Taub-NUT manifold has an exact $U(1)$ isometry that would correspond to a
rotational symmetry about the axis through the center of the two
monopoles.   In the actual solutions, the interaction between the
monopole cores gives rise to terms that break this symmetry, although
their magnitude falls exponentially with the monopole separation 
\cite{atiyah}.  

The long-range forces between the two $SU(3)$ monopoles are similar to
those in the $SU(2)$ case, except that the product of
charges (magnetic, electric, or scalar) which enters each of these
forces gives a factor of ${\bf \alpha \cdot \beta} = - {\bf \alpha}^2/2$. 
Because of this negative sign, ${\cal M}_0 $ asymptotically approaches 
a nonsingular Taub-NUT metric with positive length parameter.  In this
case, however, symmetry considerations do not exclude the possibility
of an exact identity, since the $SU(3)$ solutions with separated
monopoles have an exact axial symmetry, a property that can be traced to
the presence of the second unbroken $U(1)$. Furthermore, the Taub-NUT manifold 
with positive length parameter has a fixed point $r=0$ under its 
rotational $SU(2)$ symmetry. That could plausibly correspond to
the spherically symmetric $\gamma^*$ monopole. Note in particular that
the zero mode structure about this solution agrees precisely with what 
one would expect from the Taub-NUT metric in the neighborhood of its 
fixed point \cite{klee}.

To verify that ${\cal M}_0$ is in fact a Taub-NUT manifold, we use the
fact that the moduli space metric must be hyperk\"ahler \cite{atiyah}, so
that ${\cal M}_0 $ must be a four-dimensional manifold with a rotationally
invariant anti-self-dual metric \cite{jg}.
Any such metric can be written in the form \cite{gibbons}
\begin{equation}
g_0=f(r)^2\,dr^2+a(r)^2\sigma_1^2+b(r)^2\sigma_2^2+c(r)^2\sigma_3^2,
\end{equation}
where the line elements $\sigma_i$'s obey
$d\sigma_i=\frac{1}{2}\epsilon_{ijk}\sigma_j\wedge \sigma_k$, while
\begin{equation}
\frac{2bc}{f}\frac{da}{dr}=b^2+c^2-a^2-2\lambda bc,\quad \hbox{and cyclic
permutations thereof},
\end{equation}
where either $\lambda=1$ or $\lambda=0$. The case of $\lambda=1$ was studied 
in detail by Atiyah and Hitchin, who showed that the only three 
possibilities (up to irrelevant permutations of $\sigma_k$'s) are:

\vskip 5mm
\noindent
1) $a=b=c$ case: the flat $R^4$.

\noindent
2) $a=b\neq c$ case: the Taub-NUT geometry with positive length parameter.

\noindent
3) $a\neq b\neq c$ case: the Atiyah-Hitchin geometry
\vskip 5mm

\noindent
The case of $\lambda=0$ is even simpler and leads to only one new possibility.
\vskip 5mm
\noindent
4) the Eguchi-Hanson gravitational instanton \cite{eguchi}.
\vskip 5mm
\noindent
Only the second of these approaches the correct asymptotic geometry. Therefore,
${\cal M}_0$ must be given by the smooth Taub-NUT geometry:
\begin{equation}
g^{(4)}=(1+\frac{2l}{r})\,(dr^2+r^2\sigma_1^2+r^2\sigma_2^2)+
\frac{4l^2\sigma_3^2}{1+2l/r},
\end{equation}
where the length parameter $l$ is positive and inversely 
proportional to the reduced mass of the two monopoles. 

With an appropriate choice of the angular coordinates, the 
sum $\sigma_1^2+\sigma_2^2$ is the line element of a two-sphere $d\theta^2
+\sin^2\theta\,d\phi^2$ and is independent of the third angular coordinate
which we call $\psi$. The vector field $\xi_3=\partial/\partial\psi$ dual
to $\sigma_3$ thus become a Killing vector field and in fact generates the 
extra axial symmetry alluded to earlier. 
The ranges of $\theta$ and $\phi$ are $[0,\pi]$ and $[0,2\pi]$ respectively,
while the range of $\psi$ depends on whether the rotational isometry is
$SU(2)$ or $SO(3)$. A careful comparison with the long range interactions 
of the two monopole reveals that the conserved momentum $P_\psi$ is simply 
the $U(1)$ generator $(\alpha\cdot H-\beta\cdot H)/3$ which admits 
half-integer eigenvalues, implying that $\psi\in [0,4\pi]$. The full isometry
group of ${\cal M}_0$ is thus $SU(2)\times U(1)$.

How do we determine the discrete subgroup $\cal G$? It is most instructive
to consider the Hamiltonian ${\cal H}$
of the two monopoles at large separations. In
particular we want to concentrate on the internal motions that generate
electric charges $Q_\alpha$ and $Q_\beta$ of the respective monopoles 
\cite{klee}:
\begin{equation}
{\cal H}=\frac{1}{2}m_\alpha Q_\alpha^2+\frac{1}{2}m_\beta Q_\beta^2+
{\cal H}_{\rm rel}\left(\frac{Q_\alpha -Q_\beta}{2};\,r\right),
\end{equation}
where ${\cal H}_{\rm rel}$ encodes interaction terms that vanish at large
separation, while $m_\alpha$ and $m_\beta$ are the masses of the respective 
monopoles of magnetic charge $\alpha^*$ and $\beta^*$. The  electric charges 
are integer-quantized and thus can be realized as the canonical momenta of 
cyclic variables $\xi_\alpha$ and $\xi_\beta$ of period $2\pi$, which are
coordinates in the moduli space $\cal M$.

Note that the interaction Hamiltonian ${\cal H}_{\rm rel}$ depends only on 
the ``relative'' electric charge $Q_\alpha -Q_\beta$. Rearranging the 
Hamiltonian to factor out the noninteracting part, we find
\begin{equation}
{\cal H}=\frac{1}{2}(m_\alpha+m_\beta)\left(\frac{m_\alpha Q_\alpha+m_\beta
Q_\beta}{m_\alpha+m_\beta}\right)^2+2\frac{m_\alpha m_\beta}{m_\alpha+m_\beta}
\left(\frac{Q_\alpha-Q_\beta}{2}\right)^2+
{\cal H}_{\rm rel}\left(\frac{Q_\alpha -Q_\beta}{2};\,r\right).
\end{equation}
Now it is easy to see that we must identify the ``total'' and 
``relative'' charges with the conjugate momenta of the coordinate $\chi$ and
$\psi$ introduced above, 
\begin{equation}
P_\chi =\frac{m_\alpha Q_\alpha+m_\beta
Q_\beta}{m_\alpha+m_\beta},\qquad
P_\psi =\frac{Q_\alpha-Q_\beta}{2}.
\end{equation}
Accordingly,  $\chi$ and $\psi$ are related to the $\xi$'s by
\begin{equation}
\chi=\xi_\alpha+\xi_\beta,\qquad\psi=2\frac{m_\beta\xi_\alpha-
m_\alpha\xi_\beta}{m_\alpha+m_\beta}.
\end{equation}
Since the $\xi$'s are periodic in $2\pi$, the following identifications on
$\chi$ and $\psi$ are necessary,
\begin{equation}
(\chi,\psi)=(\chi,\psi+4\pi),\qquad
(\chi,\psi)=(\chi+2\pi,\psi+\frac{4m_\beta}{m_\alpha+m_\beta}\pi).
\end{equation}
The first simply reasserts the fact that $\psi$ has period $4\pi$, while
the second generates the discrete group ${\cal G}=Z$. For equal masses
$m_\alpha=m_\beta$, in particular, $\chi$ is also periodic in $4\pi$ 
so that the moduli space can be written in the form,
\begin{equation}
{\cal M}=R^3\times\frac{S^1\times {\cal M}_0}{Z_2}.
\end{equation}
But unless the mass-ratio is rational, the $R^1$ factor need not collapse 
to a circle by the action of the group ${\cal G}=Z$, and we have in general,
\begin{equation}
{\cal M}=R^3\times\frac{R^1\times {\cal M}_0}{Z}.
\end{equation}

\vskip 5mm 
Once the moduli space is known, the quantization of the low energy dynamics
is straightforward. In fact, for the purpose we have in mind, we will not 
even need to write down the explicit form of the Lagrangian. From the 
hyperk\"ahler property of the moduli space and the counting of fermionic
zero modes, we already know that the quantum mechanics on $\cal M$
must be  $N=4$ supersymmetric with sixteen fermionic coordinates \cite{blum}.
According to  Witten \cite{witten}, then, all zero energy states of this 
theory are in one-to-one correspondence with harmonic forms on the moduli
space.

On the other hand, there must be one and only one zero energy bound state of 
the two fundamental monopoles that carries no electric charge. This means that
the  duality hypothesis checks out only if there is a unique normalizable 
harmonic form on $(R^1\times{\cal M}_0)/Z$ invariant
under the isometries induced by $\partial/\partial\chi$ and $\partial
/\partial\psi$. The first condition can be achieved trivially by requiring
that the harmonic form does not depend on $\chi$ at all.
Furthermore since a normalizable harmonic $p$-form on the $k$-dimensional
manifold  produces a normalizable harmonic $(k-p)$-form via the 
Hodge dual procedure, the unique bound state must come from either a 
self-dual or an anti-self-dual $2$-form on ${\cal M}_0$. (This fact was 
previously emphasized by Sen \cite{sen}.)

In terms of four orthonormal $1$-forms, 
\begin{equation}
w^0= -\sqrt{1+\frac{2l}{r}}\,dr, \quad w^1=r\sqrt{1+\frac{2l}{r}}\,\sigma_1,
\quad w^2=r\sqrt{1+\frac{2l}{r}}\,\sigma_2,\quad
w^3= \frac{2l}{\sqrt{1+2l/r}}\,\sigma_3,
\end{equation}
the most general self-dual and anti-self-dual $2$-forms can be written as
\begin{equation}
F^{(\pm)}_i\,\left(2w^0\wedge  w^i \pm \epsilon_{ijk}w^j \wedge  w^k\right)
\end{equation}
Requiring that the exterior derivative vanishes, which is sufficient
for showing that they are harmonic forms, tells us that $F^{(\pm)}_i$ 
are functions only of $r$ obeying a first order differential equation 
\cite{sen}. 
It is then a straightforward exercise to show that only $F^{(-)}_3$
produces a normalizable and nonsingular harmonic $2$-form,
\begin{eqnarray}
\Omega_3^{(-)} &=& -\frac{1}{(r+2l)^2}\,\left(w^0\wedge  w^3-w^1\wedge  
w^2\right) 
\nonumber\\ &=& \frac{2l}{(r+2l)^2}\,dr\wedge 
\sigma_3+\frac{r}{r+2l}\,\sigma_1\wedge\sigma_2.
\end{eqnarray}
As noted above, this implies the existence of
a bound state of the two fundamental monopoles, which must have
the magnetic charge $\gamma^*$. This bound state is intrinsic to the
supersymmetric theory and exists only because the underlying Yang-Mills theory
has four global supersymmetries.
The $2$-form  $\Omega_3^{(-)}$ is invariant under the $U(1)$
translation generated by the Killing vector field $\partial/\partial\psi$, 
as we anticipated, not to mention being invariant under the spatial
rotation group $SU(2)$. One may regard this bound state as the
quantum analogue of the spherically symmetric $\gamma^*$ monopole found
in Ref.~\cite{erick}.

In fact, this $2$-form generates more than the purely magnetic bound 
state of two monopoles. Rather than requiring the state to be independent 
of $\chi$, we could try to construct the states with $P_\chi \neq 0$ by 
tensoring the wavefunction with $\exp{(iP_\chi\chi)}$. Since these states are
invariant under $\partial/\partial\psi$, $Q_\alpha$ is equal to $Q_\beta$,
and thus the ``total'' charge $P_\chi$ is the integer $Q_\alpha=Q_\beta$.
Furthermore, in this sector of trivial $P_\psi$, the ``total'' charge
$P_\chi$ can be identified with the eigenvalues of $\gamma\cdot H$. 
Therefore, the harmonic 
$2$-form $\Omega_3^{(-)}$ indeed generates the whole tower of dyonic states 
with magnetic charge $\gamma^*$ and integral eletric charge $n\gamma$,
as required by Montonen-Olive duality conjecture.
This provides a nice consistency check on our moduli space $\cal M$, and
specifically the division by ${\cal G}=Z$.

\vskip 5mm

Finally, we must show that $\Omega^{(-)}_3$ is indeed the unique 
normalizable harmonic form on ${\cal M}_0$. Otherwise there would a
superfluous bound state inconsistent with the duality conjecture.
For the purpose let us consider the following action of the Clifford 
algebra generated by $\Gamma^a$'s on the space of forms $\Lambda^*(
{\cal M}_0)$,
\begin{equation}
\Gamma^a(w^b)=w^a\wedge w^b-\delta^{ab},\quad\Gamma^a(w^b\wedge V)
=\Gamma^a(w^b)\wedge 
V-w^b\wedge \Gamma^a(V),\qquad \hbox{$V$ is an arbitrary $p$-form.}
\end{equation}
It is easy to see that this does represent the Clifford algebra with
$\{\Gamma^a,\Gamma^b\}= -2\delta^{ab}$; furthermore, it is well known that
the resulting Dirac operator on $\Lambda^*({\cal M}_0)$ is a sum of the 
natural exterior derivative $d$ and its adjoint \cite{spin}:
\begin{equation}
\Gamma^a\nabla_a=d+d^\dagger
\end{equation}
Thus a $p$-form is harmonic if and only if it is also a zero mode of such a
Dirac operator. Squaring the Dirac operator, we find
\begin{equation}
\Gamma^a\Gamma^b\nabla_a\nabla_b = -\nabla^a\nabla_a+ \frac{1}{4}
[\Gamma^a,\Gamma^b]\,[\nabla_a,\nabla_b]= -\nabla^a\nabla_a+{\cal R}.
\end{equation}
On $0$-forms, the curvature piece $\cal R$ is trivial as usual, while 
on $1$-forms $V_aw^a$, it reduces to a matrix multiplication
by the Ricci tensor, ${\cal R}V_a=R^b_a V_b$.
This also vanishes because the Taub-NUT space is Ricci-flat. On $2$-forms 
$V_{ab}w^a\wedge w^b$, after taking into account the vanishing Ricci tensor, 
the action of $\cal R$ is easily found to be
\begin{equation}
{\cal R}V_{ab}=-R_{ab}^{\ \ cd}V_{cd}.
\end{equation}
The anti-self-dual nature of the curvature tells us that this 
potential is nontrivial only when $V$ is anti-self-dual as well.
Finally, we do not need to consider $3$-forms and $4$-forms separately
since the Hodge dual procedure maps all normalizable harmonic $p$-forms to
$(4-p)$-forms bijectively. Therefore, with the possible exception of
anti-self-dual $2$-forms, all harmonic forms on ${\cal M}_0$ must be 
zero modes of the covariant Laplacian $-\nabla^a\nabla_a$.

Suppose a $p$-form $V$ is a zero mode of this Laplacian. By taking the 
inner product of $V$ and $-\nabla^a\nabla_a V$ and equating the result
to zero, we find
\begin{equation}
0=-\int\, \langle V,\nabla^2 V\rangle =
\int \,\langle\nabla V,\nabla V\rangle-\oint\; \langle V,\nabla V\rangle
\end{equation}
where $\langle \cdot,\cdot\rangle$ is the natural pointwise inner product of 
tensors, defined by the metric. Since $\nabla V\equiv 0$ implies a constant
$\langle V,V\rangle$, leading to either an infinite norm or an identically 
vanishing $V$, a normalizable and nontrivial solution requires that
the boundary term be nonzero. The boundary volume element 
grows as $r^2$ at the asymptotic infinity, and at least one orthonormal 
component of $V$ must have tail vanishing no faster than $1/r$.
Such an asymptotic behavior again leads to a divergent norm of $V$ 
itself so the solution is still unacceptable. 
Therefore $\Omega^{(-)}_3$ is the unique normalizable harmonic form
on ${\cal M}_0$.

The previous analysis also explains why the normalizable harmonic form 
$\Omega^{(-)}_3$ exists. The potential is diagonalized with respect to the 
decomposition of an arbitrary anti-self-dual $2$-form $V$ into three pieces 
$V^{(-)}_i =(2 V_{0i}- \epsilon_{ijk} V_{jk})$, and the diagonal entries of 
the potential, $U^{(-)}_i=-R_{0i0i}$ (no summation on $i$), are found to be 
\begin{equation}
U_1^{(-)}=U_2^{(-)}=+\frac{l}{(r+2l)^3},\qquad U_3^{(-)}=
-\frac{2l}{(r+2l)^3}.
\end{equation}
The potential is attractive only along $V^{(-)}_3$ direction precisely
where we discovered the harmonic form $\Omega^{(-)}_3$ earlier. 

\vskip 5mm
These results can be immediately extended to any group $G$ of rank $n$ 
broken to $U(1)^n$. If the $n$ preferred simple positive roots are 
${\alpha_i,i=1,...,n}$, the moduli space of a pair of fundamental monopoles 
with charges $\alpha^*_i$ and $\alpha_j^*$ is again determined by the smooth 
Taub-NUT geometry whenever $\alpha_i^*\cdot \alpha_j^* <0$, although the
action of the discrete group $\cal G$ could be modified if  Lie algebra
of $G$ is not simply-laced
\cite{klee}. The bound states of the total magnetic charge $\alpha_i^* + 
\alpha_{j}^*$ should follow similarly from our construction.

Furthermore, we believe that the moduli space of $m \: (\le n)$ distinct
fundamental monopoles should be given by a simple and tractable 
generalization of the $m=2$ case studied here. We have seen that the moduli 
space of two fundamental monopoles is obtained from its long distance behavior
alone; it is very likely that the same is true of the  moduli space of these
$m$ monopoles. The asymptotic form of this space can be found by a 
procedure analogous to that of Ref.~\cite{gary}.

This leads us to conjecture that the metric of the moduli space of our $m$
distinct fundamental monopoles is 
\begin{equation}
g^{(4m)}= M_{jk}d{\bf r}_j\cdot d{\bf r}_k+\kappa 
M^{(-1)}_{jk}(d\xi_j+W_j)(d\xi_k+W_k)
\end{equation}
where ${\bf r}_j$ is the Cartesian coordinate of the $j$-th monopole and 
$W_j$ is a weighted sum of $U(1)$ vector potential $1$-forms due to other
monopoles, i.e., $W_j=\sum_{k\neq j}\alpha_k^*\cdot\alpha_j^*{\bf A}_{k}^{(j)}
\cdot(d{\bf r}_{k}- d{\bf r}_j)$ with ${\bf A}_k^{(j)}$ being the Dirac 
potential of the $k$-th monopole evaluated at ${\bf r}_j$. Also $\kappa=
g^4/(4\pi)^2$ where we restored the magnetic coupling $g$ explicitly. 
The coordinate $\xi_j$ plays a role similar to those of $\xi_\alpha$ and 
$\xi_\beta$ above, and has periodicity $2\pi/\alpha_j^2$. Note that we have 
not factored out the center-of-mass 
coordinate. Some of the long range interaction is encoded in
the $m\times m$ matrix $M_{jk}$ which must has the form
\begin{equation}
M_{jj}=m_j-\sum_{k\neq j}\frac{g^2\alpha_k^*\cdot\alpha_j^*}{4\pi \vert 
{\bf r}_j-{\bf r}_{k}\vert},\qquad
M_{jk}=M_{kj}=\frac{g^2\alpha_k^*\cdot\alpha_j^*}{4\pi \vert {\bf r}_j-
{\bf r}_{k}\vert}, \quad j\neq k,
\end{equation}
where $m_j$'s are the BPS masses of the monopoles. It is important to note 
that we require all monopoles to be fundamental and distinct from
one another so that $\alpha_i^*\cdot\alpha_j^*\le 0$ whenever $i\neq j$.
The simplest case, $m=2$ with $(\alpha_1\cdot \alpha_2)^2/\alpha_1^2\alpha_2^2
=1/4$, is easily shown to reproduce our moduli space $\cal M$. 

While we have not shown that this is the right metric for the moduli space 
at all separations, there are some indications that it is: Apart from 
being hyperk\"ahler, this geometry has the right symmetry properties. 
The overall rotation of the $m$ monopole configuration is still an isometry 
since the metric coefficients $M_{jk}$ are dependent only on the relative 
distances. The $m$ axial rotations of $\xi_j \rightarrow \xi_j+const$ are
exactly those expected from the surviving $U(1)^n$ gauge group. Finally the 
geometry is smooth wherever two of ${\bf r}_j$ coincide.

If this conjecture is correct,
it should be a matter of straightforward, albeit
tedious, algebra to find the appropriate (anti)-self-dual $2(m-1)$-form
on the analog of ${\cal M}_0$ that represents the unique magnetic bound 
state of charge $\sum \alpha_j^*$ and the whole dyonic 
tower thereof. This problem is left for a future study.

\vskip 5mm
We thank S.J. Rey and J. Gauntlett for useful conversations. This work is 
supported in part by the U.S. Department of Energy. K.L. is supported in part 
by the NSF Presidential Young Investigator program.

\vskip 5mm 
As this work was completed, we learned of two related papers, one by 
Gauntlett and Lowe and another by Connell \cite{gaunt}.


\begin{thebibliography}{99}



\bibitem{dual}{C. Montonen and D. Olive, Phys. Lett. {\bf B 72} (1977)
117; H. Osborn, Phys. Lett. {\bf B 83} (1979) 321.}
\bibitem{sen}{A. Sen, Phys. Lett.  {\bf B 329} (1994) 217.}
\bibitem{sethi}{ S. Sethi, M.
Stern and E. Zaslow,   Nucl. Phys. {\bf B 457} (1995) 484; J.P.
Gauntlett and J. Harvey, {\it S-Duality and the Dyon Spectrum in N=2
Super Yang-Mills Theory}, hep-th/9508156.} 
\bibitem{porrati}{M. Porrati, {\it On the Existence of
States Saturating the Bogomol'nyi Bound in N=4 Supersymmetry},
hep-th/9505187.} 
\bibitem{erick}{E.J. Weinberg, Nucl. Phys.  {\bf B 167} (1980) 500.}
\bibitem{bps}{E.B. Bogomol'nyi, Sov. J. Nucl. Phys. 24 (1976) 449; M.K. Prasad
and C.M. Sommerfeld, Phys. Rev. Lett. {\bf 35} (1975) 760.} 
\bibitem{atiyah}{M.F. Atiyah and N.J. Hitchin, {\it The Geometry and 
Dynamics of Magnetic Monopoles}, Princeton Univ. Press, Princeton (1988);
Phys. Lett. {\bf A 107} (1985) 21; Phil. Trans. R. Soc. Lon. {\bf A 315}
(1985) 459.}
\bibitem{klee}{K. Lee, E.J. Weinberg and P. Yi, in preparation.} 
\bibitem{manton}{N.S. Manton, Phys. Lett. {\bf B 154} (1985) 397; (E) {\bf
B 157} (1985) 475; G.W. Gibbons and N.S. Manton, Nucl. Phys. 
{\bf B 274} (1986) 183.}
\bibitem{jg}{We thank J. Gauntlett for an informative discussion on 
four-dimensional hyperk\"ahler manifolds.}
\bibitem{gibbons}{G.W. Gibbons and C.N. Pope, Commun. Math. Phys. {\bf
66} (1979) 267.}
\bibitem{eguchi}{ T. Eguchi and A.J. Hanson, Ann. Phys. {\bf 120} (1979)
82.}

\bibitem{blum}{J.D. Blum, Phys. Lett.  {\bf B 333} (1994) 92;
J. Gauntlett, Nucl. Phys. {\bf B 411}, (1994) 443.} 
\bibitem{witten}{E. Witten, Nucl. Phys. {\bf B 202} (1982) 253.}
\bibitem{spin}{H.B. Lawson and M.-L. Michelsohn, {\it Spin Geometry},
Princeton Univ. Press, Princeton (1989).}

\bibitem{gary}{G.W. Gibbons and N.S. Manton, Phys. Lett. {\bf B 356} (1995)
32.}

\bibitem{gaunt}{J.P. Gauntlett and D.A. Lowe, {\it Dyons and S-Duality in N=4
Supersymmetric Gauge Theory}, hep-th/9601085; S.A. Connell, {\it
The dynamics of the $SU(3)$ charge (1,1) magnetic monopoles,} University
of South Australia preprint.}


\end{thebibliography}
\end{document}